\documentclass[aps,prl,amsmath,amssymb,twocolumn,superscriptaddress]{revtex4-2} 

\usepackage{graphicx}
\usepackage{dcolumn}
\usepackage{bm}
\usepackage{lipsum}

\begin{document}

\title{Generative diffusion model for surface structure discovery}
\author{Nikolaj Rønne}
\affiliation{Center for Interstellar Catalysis, Department of Physics and Astronomy, Aarhus University, DK-8000 Aarhus, Denmark}
\author{Al\'an Aspuru-Guzik}
\affiliation{Department of Chemistry, University of Toronto, Toronto, Ontario, M5S 3H6, Canada.}
\affiliation{Department of Computer Science, University of Toronto, Toronto, Ontario, M5S 2E4, Canada.}
\affiliation{Vector Institute for Artificial Intelligence, Toronto, Ontario, M5G 1M1, Canada.}
\affiliation{Acceleration Consortium, Toronto, Ontario, M5S 3H6, Canada}
\affiliation{Canadian Institute for Advanced Research, Toronto, Ontario, M5G 1M1, Canada}
\author{Bjørk Hammer}
\affiliation{Center for Interstellar Catalysis, Department of Physics and Astronomy, Aarhus University, DK-8000 Aarhus, Denmark}
\email{hammer@phys.au.dk}

\date{\today}

\begin{abstract}
We present a generative diffusion model specifically tailored to the discovery of
surface structures. The generative model takes into account
substrate registry and periodicity by including masked
atoms and $z$-directional confinement. Using a rotational
equivariant neural network architecture, we design a method that
trains a denoiser-network for diffusion alongside a
force-field for guided sampling of low-energy surface phases. An
effective data-augmentation scheme for training the denoiser-network is
proposed to scale generation far beyond structure sizes represented in
the training data.
We showcase the generative model by investigating multiple surface
systems and propose an atomistic structure model for a previously unknown
silver-oxide domain-boundary of unprecedented size. 

\end{abstract}

\maketitle

The discovery of new stable and functional materials is a prerequisite
in future technological advancements in fields such as catalysis, drug 
discovery, carbon-capture and energy storage.\cite{Yao2023}
The design-space of
stable atomistic structures is governed by quantum mechanics and it is
both vast and highly complex. Generative diffusion models have proven
highly capable of tackling the similarly vast design-space of images
and are therefore proposed to be a natural extension to ab-initio methods for
materials discovery.

The use of generative modeling in materials science has seen rapid
development in recent years by utilizing variational autoencoders\cite{gomez-bombarelli2018,lim2018,liu2018,jin2018,noh2019,yao2021},
 generative
adversarial networks\cite{kim2020,long2021,zhao2021,alverson2024},
autoregressive models\cite{gebauer2019,shi2020,westermayr2023a},
reinforcement learning\cite{jorgensen2019,simm2020,meldgaard2021,yoon2021,thiede2022,barrett2024},
language models\cite{flam-shepherd2022,fu2023,antunes2024}, normalizing flows\cite{zang2020,kuznetsov2021,ma2021,kohler2023} and
diffusion models\cite{xie2022,hoogeboom2022,lyngby2022,fung2022,jing2023,zhao2023,weiss2023,zeni2023,moustafa2023,fu2023a,cheng2023,elijosius2024a}.
Most developments have been in the regime of 
\textit{molecular} materials with fewer generative methods tackling
the more complicated case of \textit{periodic} materials. Diffusion
models lend themselves especially well to atomistic modelling due to
the few architectural requirements as compared to other
methods.\cite{song2019} Previous work on generative diffusion modeling of periodic materials has 
focused either on foundation models for discovery of bulk
structures\cite{xie2022,fung2022,zhao2023,zeni2023} or
discovery of mono-layer materials with no
support\cite{lyngby2022,moustafa2023}. The seminal work by Xie et
al.\cite{xie2022}, tackling bulk structure generation, uses a diffusion
model as the decoder in a variational autoencoder to address the
inversion problem in materials discovery. A prerequisite for these
developments in generative modeling of periodic materials is
the advent of equivariant machine learning force-fields
(MLFF)\cite{thomas2018,schutt2021,gasteiger2021,batzner2022}.
These network architectures are able to fully capture the symmetries
of periodic materials in 3D and open the possibility to design
equivariant score-based diffusion models by utilizing the vectorial 
atom embeddings to predict vectorial scores. 

In this work, we leverage the new developments in generative modelling
and naturally extend it to surface structure discovery. We introduce
a generative diffusion method specifically tailored to surface
supported thin-films by including masked 
substrate atoms and a $z$-directional confinement. Furthermore, we
introduce a force-field-guidance framework 
for sampling by introducing learned atomic forces into the sampling
process and an efficient data-augmentation technique to extent model
stability to larger systems.
We show the superiority of the diffusion based generation method by
comparison to random structure search\cite{pickard2011} (RSS) for multiple
surface systems. Furthermore, we solve a domain-boundary problem for a
silver-oxide on Ag$(111)$. 

\begin{figure}
  \includegraphics{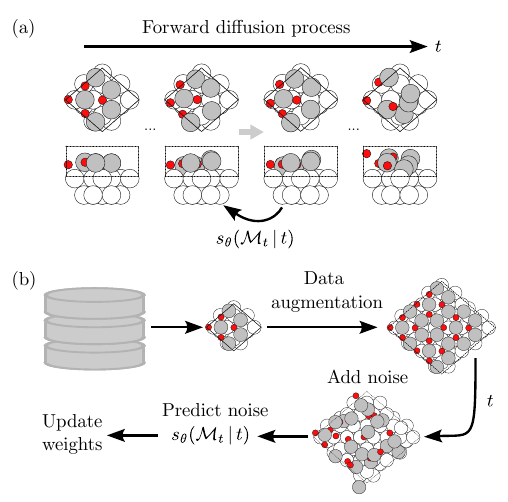}
\caption{\label{fig:schematic} (a) Schematic of the forward diffusion
  process where a small amount of noise is added at each step. At
  the final step the atomic positions correspond to being drawn from a
  uniform distribution. The denoiser-network
  , $s_\theta(\mathcal{M}\, | \, t)$, learns to predict the
  added noise at any given time-step $t$. (b) The process for
  training the denoiser-network. A training step is comprised of
  drawing a sample, performing data-augmention through periodic
  repetition of sample, drawing a random time $t\in[0,1]$, 
  corrupt the sample corresponding to $t$, predict the added noise
  using the denoiser-network and finally use the disagreement
  between added and predicted noise to update the denoiser-network weights. 
}
\end{figure}

Generative diffusion models rely on
learning to denoise deliberately corrupted data hereby establishing a
denoiser-network. The trained denoiser-network can then be used to
sample from the underlying data distribution. Both the forward data-corruption
process and the reversed generation process can be formulated as
stochastic differential equations (SDE). The denoiser-network
naturally appears in the reversed SDE as the score function, $s$,
defined as the derivative of the log-transformed data probability
distribution.\cite{song2021}

A periodic atomistic structure is
fully described by $\mathcal{M} = (\mathbf{z}, \mathbf{R},
\mathbf{S})$ representing the atomic 
types, atomic positions and periodic cell. In this work, we keep
atomic types and cell fixed through the diffusion process. We define a
forward diffusion process acting on the atomic positions through a
SDE as
\begin{equation}\label{eq:forSDE}
  d\mathcal{M} = -\frac{1}{2}\beta(t) \mathcal{M} dt + \sqrt{\beta(t)} d\mathcal{W},
\end{equation}
where $t \in [0,1]$ is time,
$\beta(t) = \beta_{\mathrm{min}} + t\left(\beta_{\mathrm{max}}-\beta_{\mathrm{min}}\right)$ is a linearly 
increasing function between $\beta_{\mathrm{min}}$ and
$\beta_{\mathrm{max}}$, and $d\mathcal{W}$ is infinitesimal Brownian
noise.\cite{song2021} The coefficients of the terms are typically
referred to as \textit{drift} and \textit{diffusion} coefficient respectively. The forward diffusion
process is analytical and can be discretized and sampled. We index the
discretized diffusion process by the time variable, which together with a known noise
distribution allows a direct forward sampling of noisy structures. Intuitively
the forward SDE represents the iterative addition of Gaussian noise to the atomic positions as illustrated in
Fig.\ \ref{fig:schematic}(a).
The reversed diffusion process can also be formulated as a SDE and is given as
\begin{equation}\label{eq:revSDE}
  d\mathcal{M} = \left[-\frac{1}{2}\beta(t)\mathcal{M} - \beta(t)
    s(\mathcal{M}) \right]dt + \sqrt{\beta(t)} d\tilde{\mathcal{W}},
\end{equation}
where $s(\mathcal{M})$ is the unknown score function and
$d\tilde{\mathcal{W}}$ is reverse-time Brownian noise.
We design a network that approximates the score function,
$s_\theta(\mathcal{M}\, |\, t)$, given  
a structure and a time conditioning $t \in [0,1]$ and will refer to
this as the denoiser-network. 
The denoiser-network, $s_\theta(\mathcal{M}\, |\, t)$, is trained by predicting the
vectorial noise added at any given time, $t$, through denoising
score-matching\cite{song2019} as illustrated in
Fig.\ \ref{fig:schematic}(b). The architecture  
of the denoiser-network is a time-conditioned non-conservative
vector-field regressor implemented using the equivariant PaiNN
\cite{schutt2021} architecture cf. SI Fig. 1 for more details. 

The denoiser-network training process is outlined schematically in
Fig.\ \ref{fig:schematic}(b) and is presented in detail in the
following. First, a database of stable atomistic structures is
procured. In this work we, either perform RSS for multiple
stoichiometries in small periodic cells and gather the lowest energy
structures, or we use a previously
established dataset. The specific composition of the dataset is highly
influential on the later generated structures as it defines the data
probability distribution that is sampled from. This acts as a
data bias, that can be used to drive generation towards specific
areas of configurational space. Specifically we select low-energy
structures as training data. A further discussion of this is
provided in the supplementary material. For all cases, we use smaller
systems for training data than the system we use for
testing, since these are feasible to evaluate with
DFT. Furthermore this ensures that we showcase out-of-distribution
generalizability of the generative model. Second, we perform data-augmentation by repeating the training
data periodically, such that the effective number of atoms in the
training data materials increases. This allows the denoiser-network to
learn interactions between a higher number of atoms at no cost of
procuring additional training data for larger cell-sizes. This 
data-augmentation technique can equally well be used for bulk
materials and ensures stability of the denoiser-network when many atoms
are close. The extent of data-augmentation needed depends on the
expected size of the test systems. Third, denoising score-matching is
used to train the denoiser network. Here, a random timestep, $t\in [0,1]$, is sampled and the training
example is noised according to the forward diffusion process. The
denoiser-network is then used to predict the noise added to 
the training example based on the noisy structure and conditioned on
$t$. The network loss is the discrepancy between the actual
noise and the predicted noise, taking periodic boundary conditions into
account, which is then used to update the network weights.\cite{xie2022} In practice
training is performed with batches of data.

We introduce two specific developments tailoring the 
diffusion model to surface structure generation. First, we use
truncated Gaussian distribution in the $z$-direction perpendicular to
the substrate surface, while we use a standard Gaussian distribution
in the periodic directions cf. SI Fig. 2. Hereby, we ensure that
generated structures only form surface structures on one side of the
finite substrate. Second, substrate atoms are masked in the diffusion process,
hereby fixing the atoms. Masking substrate atoms is key when studying
surface systems as the substrate is often experimentally known, whereas
the overlayer is what dictates the physical and chemical
properties.\cite{li2023} In previous work\cite{lyngby2022}, 
substrate interactions have been neglected, while we introduce 
a masked substrate that allows for the direct discovery of surface and
registry dependent structural models.


\begin{figure}
  \includegraphics{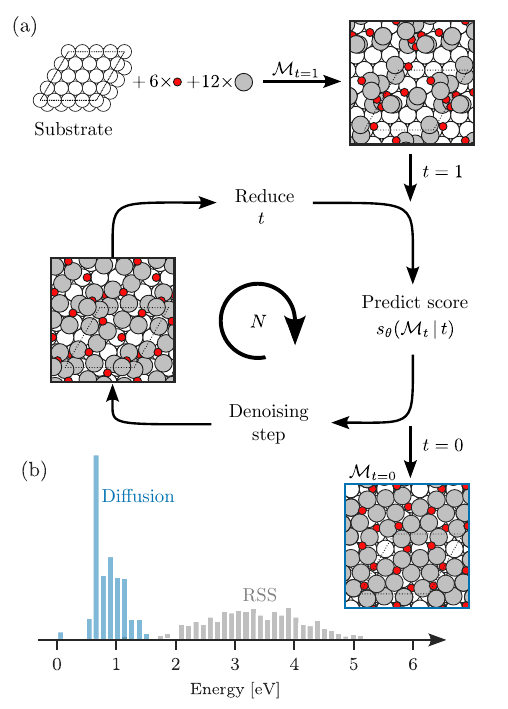}
  \caption{\label{fig:sampling}
    (a) Sampling process exemplified with the $p(4\times 4)$ AgO
    surface structure. The $t=1$ structure is initialized at
    random. $N$ denoising steps are performed following Eq.\
    \ref{eq:revSDE} using the predicted score from the
    denoiser-network and slowly annealing $t$. The $p(4\times 4)$
    structure sampled from the diffusion process is presented. (b)
    Comparison between diffusion-based sampling and RSS for
    $\mathrm{Ag}_{12}\mathrm{O}_{6}$ in the $p(4\times 4)$ cell.
}
\end{figure}

Sampling from a diffusion model consists of following the reversed
diffusion process described by Eq.\ \ref{eq:revSDE} using the trained
denoiser-network and is schematically presented in
Fig.\ \ref{fig:sampling}(a). First, the test substrate and the
pre-specified number of atoms is initialized according to the known prior
noise distribution, which is a uniform distribution of each atom
within the periodic cell. The prior noise distribution is
intuitively understood as the distribution obtained by iterative
application of Gaussian noise. We use the 
$\mathrm{Ag}(111)$-$p(4\times 4)$ cell with stoichiometry
$\mathrm{Ag}_{12}\mathrm{O}_{6}$ representing the  
cell and stoichiomtry of the known ``$\mathrm{Ag}_6$
model'' as an example in Fig.\ \ref{fig:sampling}(a). Second, the time-variable, $t \in \{1, ..., 0\}$,
is discretized in $N$ steps of step-length $\Delta t$.
Third, $N$ steps of a time-annealing process is performed where the
denoising follows Eq.\ \ref{eq:revSDE} using the predicted score as
shown in Fig.\ \ref{fig:sampling}(a).
During sampling, the same noise distributions are used
as in the forward process. Specifically in the $z$-direction, the
truncated Gaussian distribution is used such that the atoms never
leave the non-periodic confinement. 
We introduce force-field-guidance inspired by
classifier-guidance for image generation\cite{dhariwal2021}, where the
force-prediction of a concurrently trained MLFF, $f(\mathcal{M})$, is
gradually introduced during the sampling process in 
order to guide the reversed diffusion towards more stable
materials. The MLFF can be efficiently included in the denoiser
architecture almost without additional prediction cost cf. SI
Fig. 1. The force injection is controlled by the scalar function
$\tau(t)$ as 
\begin{equation}
  d\hat{\epsilon} = \tau(t) f(\mathcal{M}) dt,
\end{equation}
where $\tau (t) = \eta (1-t)$ with $\eta$ a constant and
$f(\mathcal{M}) =-\nabla_{\mathbf{R}} E(\mathcal{M})$. This means that, as time is
decreasing during sampling, more emphasis is put onto decreasing the
predicted force. $d \hat{\epsilon}$ represents an
additional term in the reversed diffusion 
process defined by Eq.\ \ref{eq:revSDE} and thus becomes an integrated
part of the denoising step. 

To judge the quality of sampled structures, we
perform a comparison between diffusion-based sampling and the unbiased
sampling obtained using RSS. We relax every sampled structure to provide a fair comparison in
terms of energy of the generated
structures. Fig.\ \ref{fig:sampling}(b) shows a typical comparison to
RSS in terms of stability of generated structures for the ``$\mathrm{Ag}_6$
model'' system in the $p(4\times 4)$ cell. In this case, the diffusion model is trained on
small $c(4\times 8)$ unit cell structures of varying
stoichiometry generated using RSS. An example of a training structure
is shown in Fig.\ \ref{fig:schematic}(a). A detailed discussion on the choice
and influence of training data is provided in the supplementary
material.

Figure \ref{fig:sampling}(b) evidences that the structures produced by sampling from
the diffusion model are lower in energy compared to what is produced
by RSS. This demonstrates that the diffusion model is capable of
generalizing to larger and more complex atomistic systems. The
data bias of the method is likewise transferable giving rise to
a shifted and more narrow energy-distribution of samples compared to RSS.
The three atomistic structural models shown as insets in
Fig.\ \ref{fig:sampling} represents the first, middle 
and final structural models during a sampling trajectory. In the middle of sampling, corresponding
to step $t=0.5$, the material is clearly more ordered with local
motifs starting to resemble chemically stable bonding patterns. The
final material represents the ``$\mathrm{Ag}_6$ model'' of
silver-oxide on Ag$(111)$ with a central ring, that exposes a
substrate atom. We observe that the surface registry is found directly
by including the masked substrate atoms in the diffusion process.

\begin{figure}
  \includegraphics{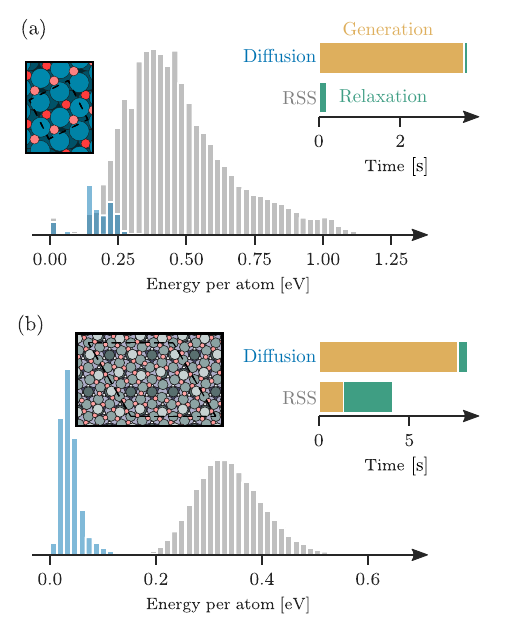}
  \caption{\label{fig:energy}
    Comparison between diffusion sampling and RSS with equal GPU-time for (a) $\mathrm{Pd}_4\mathrm{O}_4$ on
    $\mathrm{Pd}(100)$ in a $p(\sqrt{5}\times \sqrt{5})$ cell giving rise to the
    $(\sqrt{5} \times \sqrt{5})\mathrm{R}27^{\circ}$ surface oxide as
    shown in the inset. (b) $\mathrm{Sn}_{43}\mathrm{O}_{46}$ on $\mathrm{Pt}_3\mathrm{Sn}(111)$
    substrate in a $p(8\times 8)$ supercell giving rise to a defected oxide
    structure as presented in the inset. Furthermore, a timing
    comparison between diffusion and RSS detailing time spent on
    generation and relaxation per generated
    structure. 
  }
\end{figure}

In Fig.\ \ref{fig:energy} we further demonstrate that the method is generally
applicable to surface structure discovery. We concurrently compare the
quality and computational efficiency of diffusion 
and RSS. This is done by limiting the total GPU-time and
generate as many candidates as possible either using RSS or
diffusion. As a target potential, we use a pre-trained MLFF trained using the
SchNetPack software\cite{schutt2018,schutt2023}. We present timings for generation and
relaxation. In Fig.\ \ref{fig:energy}(a) we study a
rather small system of PdO on $\mathrm{Pd}(100)$ in a $p(\sqrt{5}\times \sqrt{5})$
cell giving rise to the $(\sqrt{5} \times \sqrt{5})\mathrm{R}27^{\circ}$
surface oxide as shown in the inset.\cite{kostelnik2007} Similar to the AgO system
shown previously, the structures generated by the diffusion model are
much more stable than those produced by RSS. But within the fixed GPU
time-budget we are able to produce many more samples using RSS thereby 
giving similar performance in sampling the most stable
structures. Comparing the time for generation and sampling, we see that
little relaxation is needed for diffusion generated samples,
which reflects that the diffusion-produced samples are already very
close to local energy minima. In Fig.\ \ref{fig:energy}(b) we move to
a much more complicated system consisting of a defected SnO on
$\mathrm{Pt}_3\mathrm{Sn}(111)$ substrate in a $p(8\times 8)$
supercell. We choose the stoichiometry of
$\mathrm{Sn}_{43}\mathrm{O}_{46}$ which deviates slightly from the
known $\mathrm{Sn}_{11}\mathrm{O}_{12}$ stoichiometry of the $p(4 \times 4)$ 
surface phase.\cite{merte2022} For such a complicated system it is a rare-event
that RSS generates a stable structure, whereas the diffusion model is
able to scale efficiently to larger systems and continue to generate
highly stable samples. Furthermore, the generation time relative to
RSS is not as limiting, since both generation and especially relaxation
is very time-consuming with RSS. Similar to
Fig.\ \ref{fig:energy}(a), we also see that the relaxation-time for
diffusion is significantly less than for RSS again showing that the
structures sampled from the diffusion model are close to local
energy minima. The structural model presented in
Fig.\ \ref{fig:energy}(b) is the most stable generated structure and
represents the $p(4 \times 4)$ phase with a single $\mathrm{SnO}_2$-unit
removed highly resembling the defected experimental STM images
presented in Ref.\ \cite{merte2023}.

\begin{figure}
  \includegraphics{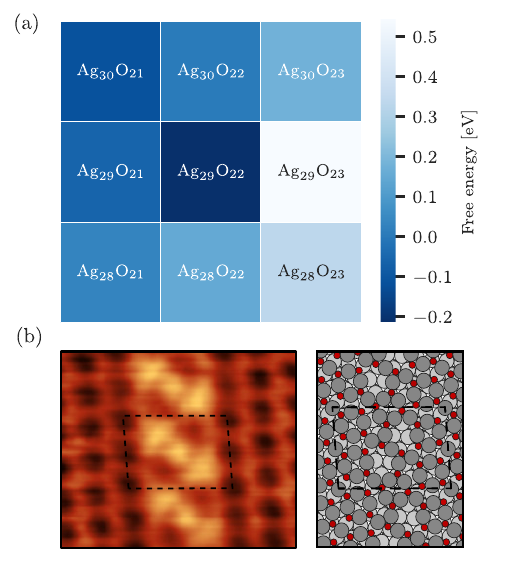}
  \caption{\label{fig:ago-defect} (a) Thermodynamic stability of best
    domain-boundary structures for varying stoichiometries under
    experimental conditions. (b) Experimental STM and proposed
    structural model of the $\mathrm{Ag}_{29}\mathrm{O}_{22}$
    domain-boundary. Experimental STM adapted with permission from
    ACS Catal. 2016, 6, 7, 4640–4646. Copyright 2016 American Chemical
    Society.}
\end{figure}

To further test the out-of-domain generalizability of diffusion sampling, we use a
diffusion model trained on stable silver oxides structures in the
small $c(4 \times 8)$ unit cell to perform a diffusion-based structure
search for a defected surface phase that has been 
experimentally observed in Ref.\ \cite{derouin2016}. The cell size of the test
system is $\sqrt{57}R6.6^\circ\times\sqrt{28}R40.9^\circ$. The screening workflow is outlined in the
following. First, we generate structures with the diffusion models for 
varying stoichiometries as indicated in Fig.\ \ref{fig:ago-defect}. Second,
we perform post-relaxation of all generated structures using the
universal potential CHGNet\cite{deng2023}. Last, we select the ten
most stable structures for relaxation with DFT and evaluate the
thermodynamic stability\cite{ronne2022}. Fig.\ \ref{fig:ago-defect} presents the
thermodynamic stability of the best surface phases under experimental
conditions in (a) and the proposed atomistic model of the surface defect together with
experimental STM hereof in (b). The proposed structural model features
a very similar motif as the $c(4 \times 8)$ phase that forms the left and
right boundary and it should be noted that all surface atoms are
generated using diffusion without forcing the boundary to be that of
the $c(4 \times 8)$ phase. This exemplifies the power of
data-driven generation through diffusion sampling and provides a major
improvement over unbiased approaches such as RSS, where
discovering the stable surface defect would be unfeasible. 

In conclusion, we have introduced a generative diffusion model for
surface structure discovery that is shown to outperform traditional
atomistic generation methods on a number of surface
systems. We have shown that a data-biased generative method, even
though more computational demanding, provides more quantum
mechanically relevant samples than a relaxation based
approach. Finally, we have proposed a structural model to a large 
silver-oxide domain-boundary based on a diffusion-based structure search.

The code is publicly available under the GPLv3 license at \url{https://github.com/nronne/dss}
and all data is available at \url{https://github.com/nronne/surface-diffusion-model-data}.

We acknowledge support by VILLUM FONDEN through Investigator grant,
project no. 16562, and by the Danish National Research Foundation
through the Center of Excellence “InterCat” (Grant agreement no:
DNRF150). This research was undertaken thanks in part to funding
provided to the University of Toronto’s Acceleration Consortium from
the Canada First Research Excellence Fund CFREF-2022-00042. AAG thanks
Anders G. Frøseth for his generous support. AAG also acknowledges the
generous support of Natural Resources Canada and the Canada 150
Research Chairs program. 

\section{References}
\bibliographystyle{apsrev4-1}
\bibliography{bib}

\end{document}